\documentstyle[11pt,newpasp,twoside,epsf]{article}

\begin{document}

\title{Radio Afterglows of Gamma-Ray Bursts: Unique Clues to the
Energetics and Environments}

\author{Edo Berger}
\affil{California Institute of Technology}

\begin{abstract}
Radio observations of gamma-ray burst (GRB) afterglows provide both
complementary and unique diagnostics of the afterglow physics and
environment of the burst.  Here we concentrate on three unique aspects
of GRB energetics and environments afforded by radio and submillimeter
observations: the non-relativistic evolution of the fireball, the
density profile of the circumburst medium, and the study of obscured
star formation in GRB host galaxies. 
\end{abstract}

\section{Introduction}

\begin{figure}[t]
\plotone{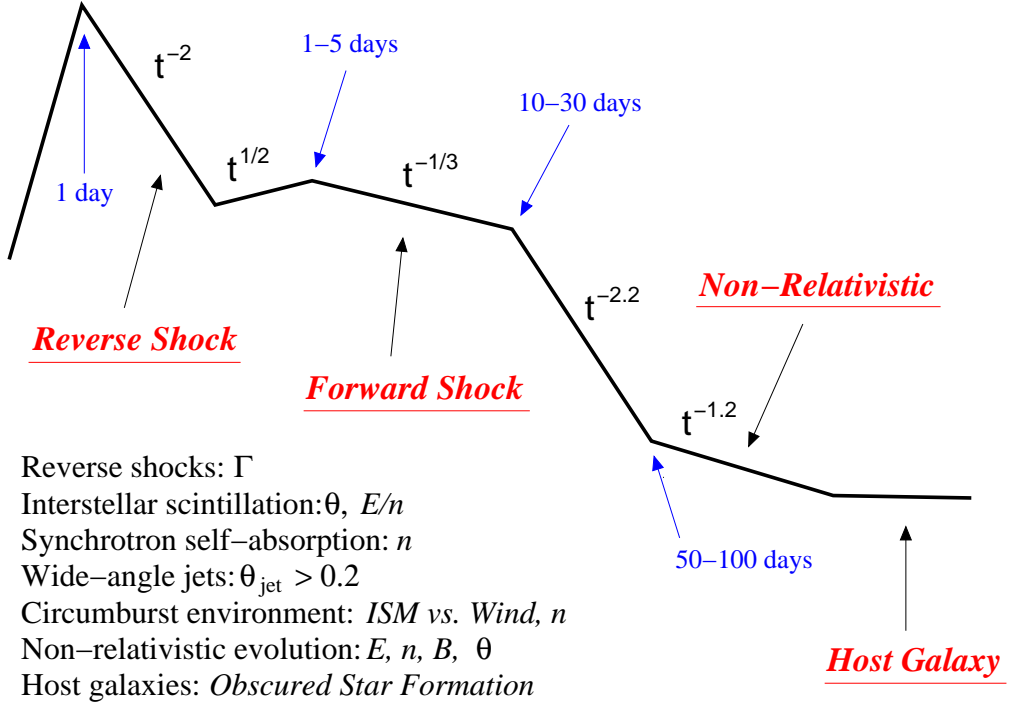}
\caption{Heuristic lightcurve in the radio band.  Timescales and
scalings for the temporal evolution are indicated.  The list
summarizes aspects of the flux evolution which are unique to the radio
bands (Lorentz factor, $\Gamma$; source size, $\theta$; energy, $E$;
density, $n$; jet opening angle, $\theta_{\rm jet}$; density profile;
magnetic field strength, $B$; and obscured star formation rate).}
\label{fig:lc}
\end{figure}

The afterglows of gamma-ray bursts (GRBs) are a broad-band
phenomenon.  As such, their study requires exquisite data from radio
to X-rays.  Still, in addition to complementary information, the radio
band provides some unique diagnostics of the afterglow physics and
burst environment.  We illustrate this in Figure~\ref{fig:lc} which 
shows a heuristic radio lightcurve ranging in time from a few hours to
hundreds of days after the burst.  Here we concentrate on three
unique aspects of radio observations: the non-relativistic expansion,
the ability to probe the density and profile of the circumburst
medium, and as a tracer of obscured star formation in GRB host
galaxies.  However, before delving into this discussion we quickly
summarize the information  in Figure~\ref{fig:lc} by way of
introduction. 

At present, even with response times to GRB alerts of minutes the
radio bands provide the best way to study the synchrotron emission
from the reverse shock (e.g.~Soderberg \& Ramirez-Ruiz 2002).  Optical
observations require response on the timescale of the burst duration
and have been successfully made only three times (Akerlof et al.~1999;
Kobayashi \& Zhang 2002; Fox et al.~2002; Li et al.~2003).  Only in
the case of GRB\,990123, the peak of the reverse shock emission was
actually observed.  In the radio bands on the other hand, emission
from the reverse shock has been observed several times since the peak
is more easily observable is $t\sim \sim t_{\rm dur}\times (\nu_{\rm
rad}/\nu_{\rm opt})^{-48/73}\sim 1$ day (Sari \& Piran 1999).  In
addition, the mere detection of the reverse shock in the radio on the
timescale of 1 day rules out a circumburst medium with a Wind
(i.e.~$\rho\propto r^{-2}$) density profile (Berger et al.~2003a).

By the same token, the typical response time prevents optical
observers from catching the peak of the synchrotron emission from the
forward shock.  Thus, optical observations cannot constrain the peak
flux which is required for inferring the physical parameters of the
burst and environment
(e.g.~Berger et al.~2000).  The radio bands directly trace these two
parameters since the peak of the spectrum reaches the radio on the
timescale of $\sim 10$ days.  Moreover, only the radio bands trace
synchrotron self-absorption, which is particularly sensitive to the
density of the circumburst medium.

Perhaps the most unique aspect of the radio emission is the existence
of propagation effects in the form of interstellar scintillation.  A
detailed discussion of scintillation is beyond the scope of this
review, but it serves as a unique way to ``resolve'' the afterglow and
infer the size of the fireball over time (e.g.~Frail, Waxman \&
Kulkarni 2000). 

Radio observations are also well-suited for inferring the opening
angles of wide jets.  The signature of a collimated outflow is a break
in the lightcurves occurring when $\Gamma\sim \theta_j^{-1}$ (Rhoads
1997); for wide jets the break occurs at late time.  Unfortunately, on
such timescales the host galaxy masks the optical afterglow and the
break cannot be inferred from this band.  In the radio, on the other
hand, the emission usually peaks on the timescale of about 10 days so
wide jets are easily studied (Berger et al.~2001).

The dominance of the optical host galaxy at late time in addition to
the faint X-ray afterglow indicate that only the radio bands allow us
to study the long-term behavior of the afterglow.  In several cases we
have observed radio afterglows over a year after the burst.  One
effect, which is readily detectable if the afterglow is bright enough,
is the transition from relativistic to sub-relativistic expansion
(Frail et al.~2000).  Depending on the initial kinetic energy of the
ejecta and the density of the ambient medium, this occurs on the
timescale of $\sim 100$ days (Figure~\ref{fig:lc} and
\S\ref{sec:nr}). 

Finally, as the radio emission fades significantly we may detect
emission from the host galaxy.  In all cases in which a radio host
has been detected (e.g.~Berger et al.~2003b) the inferred star
formation rates are several hundred of M$_\odot$ yr$^{-1}$, indicating
significant dust obscuration (\S\ref{sec:host}).

To summarize, radio observations probe the evolution of the fireball
over a factor of $\sim 10^3$ in time.  Below we focus on the
information that can be gleaned from these unique observations for the
study of GRB energetics and environments.

\begin{figure}[htbp]
\plotone{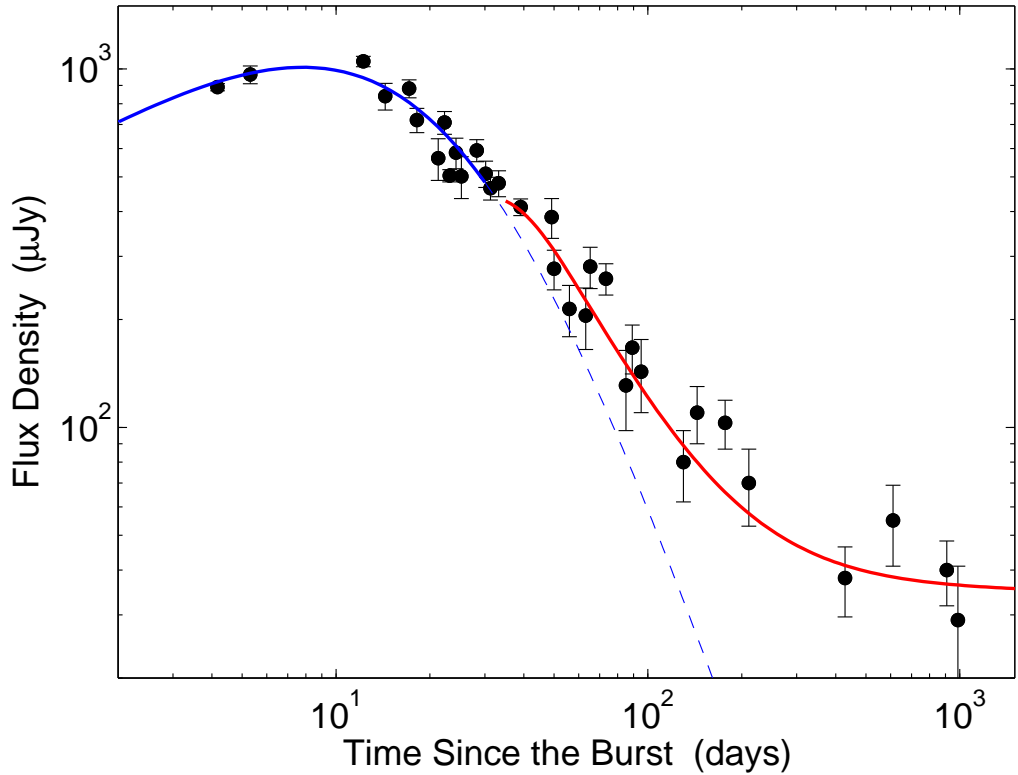}
\caption{Radio lightcurve of GRB\,980703 at 8.46 GHz.  The blue line
is a model fitting the first 50 days of data.  Clearly an
extrapolation of this model to late time (dashed blue line) does not
agree with the data.  This is due to a transition to sub-relativistic
expansion.  The red line is a model fit using the Sedov-Taylor model
with a host galaxy dominating at late time.  From Berger et al.~in
prep.} 
\label{fig:980703}
\end{figure}

\section{The Non-Relativistic Evolution: Fireball Calorimetry}
\label{sec:nr}

Recent work has shown that the beaming-corrected $\gamma$-ray and
kinetic energies of GRBs are approximately constant (Frail et
al.~2001; Berger, Kulkarni \& Frail.~2003).  The latter, however, is
inferred without knowledge of what this constant energy actually is.
Instead, the kinetic energy in the afterglow phase can be inferred
from broad-band observations (e.g.~Panaitescu \& Kumar 2002).  In
reality this requires exquisite data and assumptions about the
hydrodynamics of the fireball which are yet to be verified.

Fortunately, radio observations at late time provide an alternative
method.  The principle is simple.  As the fireball decelerates it
eventually becomes trans- and then non-relativistic.  This typically
happens on the timescale of $\sim 100$ days when the peak of the
synchrotron spectrum is in the radio bands.  A simple signature of the
transition from relativistic to sub-relativistic expansion is a
flattening of the radio lightcurves compared to the jet evolution
(Figure~\ref{fig:980703}).  Analyzing the subset of radio data
following this transition in the Sedov-Taylor self-similar framework
we can infer the energy, density, magnetic field strength, and size of
the fireball.  This method has the principle attraction that the
dynamics of the spherical, non-relativistic fireball are better
understood than those of relativistic expanding jets.

To date this method has been applied to two bursts, GRB\,970508 (Frail
et al.~2000) and GRB\,980703 (Berger et al.~in prep).  In both cases
energies of about $10^{50}$ erg have been inferred, in agreement with
other methods.  One possibility which is currently being investigated
is whether there is evidence for evolution of the fireball parameters
between the early observations and the late non-relativistic stage.

\section{The Circumburst Environment: ISM vs.~Wind}

One of the indirect clues to the identity of GRB progenitors is the
structure and density of the circumburst medium.  If the progenitors
are massive stars then we expect them to explode in the dense media of
their birth sites.  Moreover, the pre-burst stellar mass loss is
expected to influence the environment, resulting (in the most simple
case) in a density profile $\rho\propto r^{-2}$.  On
the other hand, if GRBs arise from delayed mergers of compact source
binaries then the environment is expected to be the tenuous ISM of the
host galaxy.  This is because velocity kicks imparted to the system in
the supernova explosion of the binary members will eject it away from
the birth-site.  

Extensive efforts have been made to infer the density and profile of
the circumburst medium primarily by studying the broad-band
afterglow emission (e.g.~Chevalier \& Li 2000).  Unfortunately,
these studies have been inconclusive in distinguishing between Wind
and ISM density profiless due primarily to the lack of early radio and
submillimeter observations, or in some cases their preclusion from the
analysis.

There are two clear exceptions to this disappointing trend, which in 
conjunction present an interesting conundrum.  In the case of
GRB\,011121 Price et al.~(2002) inferred a Wind medium thanks to
dual-band radio observations.  At the same time, Bloom et al.~(2002)
interpreted the late-time red bump in the optical afterglow as a
supernova that exploded at the same time as the burst.  These
observations naturally point to a massive star as the progenitor of
this burst.  On the other hand, for GRB\,020405, Berger et al.~(2003a)
showed that the reverse shock emission detected at $t\sim 1$ day in
the radio bands rules out a Wind medium.  However, this burst also had
an associated red bump indicative of a supernova (Price et al.~2003)
and hence a massive stellar progenitor.

These results imply that contrary to the optimistic expectations of
afterglow modelers, a Wind profile does not necessarily accompany a
massive progenitor and may not be a strong clue to the nature of the 
progenitor.  Still, the inference of Wind vs.~ISM profile relies
heavily on radio observations.

\section{The Large-Scale Environment: Host Galaxies and Obscured Star
Formation}
\label{sec:host}

The host galaxies of GRBs can be used as a complementary and perhaps
quite promising sample for the study of cosmic star formation: \\
{\it Redshifts:} Thanks to the bright afterglows, the redshift of the
host galaxy can be determined regardless of its brightness, via
absorption spectroscopy. \\
{\it Faint (Dwarf) Hosts:} Some GRB host galaxies are several
orders of magnitude fainter than $L_*$ and provide us with a glimpse
of faint dwarfs at high redshift. \\
{\it Immunity to Dust:} The dust-penetrating power of GRBs results in
a sample that is independent of the dust properties of the individual 
galaxies. \\
{\it Very High Redshifts:} GRBs are detectable to $z>10$ (should they
exist). 
\begin{figure}[t]
\plottwo{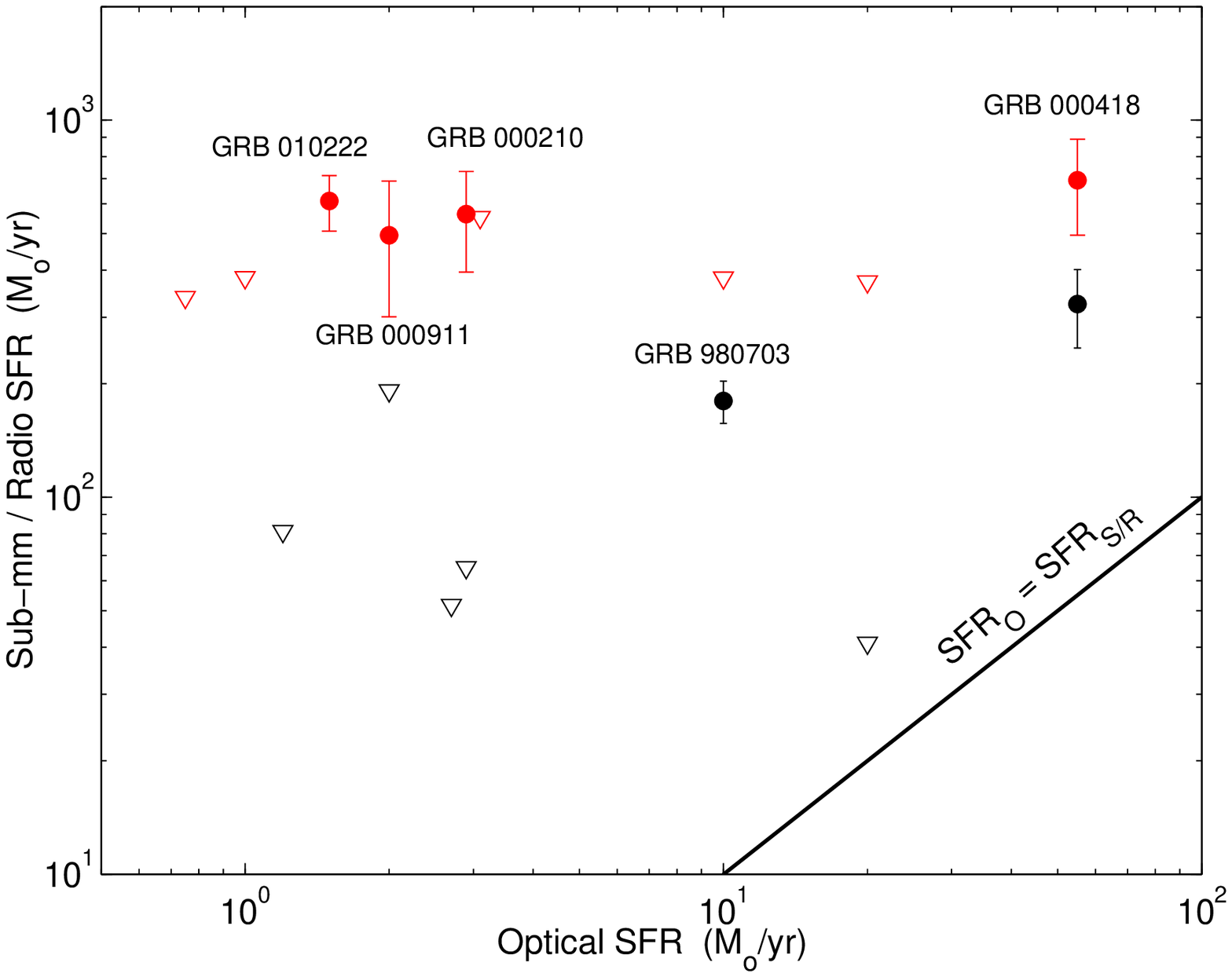}{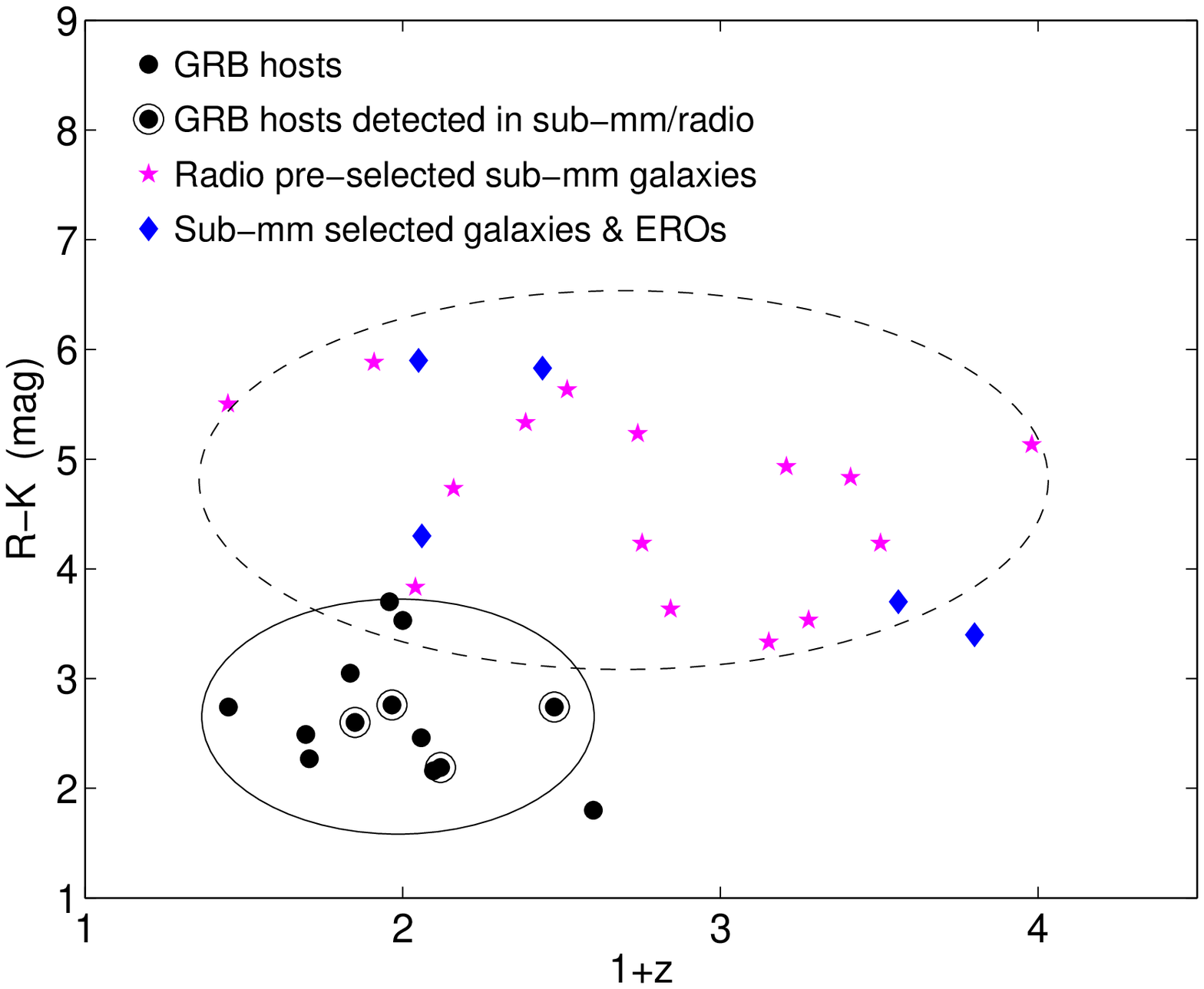}
\caption{{\it Left:} Submillimeter/radio vs.~optical star formation
rates for GRB host galaxies.  The line designates a one-to-one
correspondence between the two SFR estimates.  Clearly, some hosts
have a high fraction of obscured star formation.  {\it Right:} $R-K$
color as a function of redshift.  The ellipses are centered on the
mean color and redshift for each population of galaxies and have
widths of $2\sigma$.  The GRB hosts are in general significantly bluer
than submillimeter galaxies in the same redshift range.}
\label{fig:sfr}
\end{figure}
 
To date, GRB host galaxies have mainly been studied in the optical and
NIR bands.  With the exception of one source (GRB\,020124; Berger et
al.~2002), every GRB localized to a sub-arcsecond position has 
been associated with a star-forming galaxy, with star formation rates
of $\sim 1-10$ M$_\odot$ yr$^{-1}$.  

On the other hand, recent radio and submillimeter observations
(Berger, Kulkarni \& Frail 2001; Frail et al.~2002; Berger et
al.~2003b) have revealed that about 20\% of GRB hosts are
ultra-luminous with star formation rates of several hundred of
M$_\odot$ yr$^{-1}$, indicating significant dust obscuration
(Figure~\ref{fig:sfr}).    
While these properties are similar to those of submillimeter-selected
galaxies (Chapman et al.~2002) the $R-K$ colors of GRB hosts are
significantly bluer (Figure~\ref{fig:sfr}).  Exactly what this means is
still not clear (one possibility is younger starbursts), but it is
evident that GRBs probe a portion of the galaxy population that is
completely missed in current submillimeter surveys.   

While the present GRB sample is significantly smaller than the
Lyman-break or submillimeter samples, SWIFT should supply us
with several hundred GRB host galaxies.  This sample may provide a new
window to the evolution of cosmic star formation.

\acknowledgments
We wish to thank numerous collaborators, in particular Dale Frail and
Shri Kulkarni, and support from NSF and NASA grants.


\begin{references}
\reference Akerlof, C.~W., et al. 1999, Nature, 398, 400
\reference Berger, E., et al. 2000, ApJ, 545, 56
\reference Berger, E., et al. 2001, ApJ, 556, 556
\reference Berger, E., Kulkarni, S.~R. \& Frail, D.~A. 2001, ApJ, 560,
652
\reference Berger, E., et al. 2002, ApJ, 581, 981
\reference Berger, E., et al. 2003a, ApJ, 587, L5 
\reference Berger, E., et al. 2003b, ApJ in press; astro-ph/0210645
\reference Berger, E., Kulkarni, S.~R. \& Frail, D.~A. 2003, ApJ in
press; astro-ph/0301268
\reference Bloom, J.~S., et al. 2002, ApJ, 572, L45
\reference Chevalier, R.~A. \& Li, Z. 2000, ApJ, 536, 195
\reference Fox, D.~W., et al. 2002, Nature, 422, 284
\reference Frail, D.~A., Waxman, E. \& Kulkarni, S.~R. 2000, ApJ, 537,
191
\reference Frail, D.~A., et al. 2001, ApJ, 562, L55
\reference Frail, D.~A., et al. 2002, ApJ, 565, 829
\reference Kobayashi,S. \& Zhang, B. 2002, ApJ, 582, L75
\reference Li, W., et al. 2003, ApJ, 586, L9
\reference Panaitescu, A. \& Kumar, P. 2002, ApJ, 571, 779
\reference Price, P.~A., et al. 2002, ApJ, 572, L51
\reference Price, P.~A., et al. 2003, ApJ in press; astro-ph/0208008
\reference Rhoads, J.~E. 1997, ApJ, 487, L1
\reference Sari, R. \& Piran, T. 1999, ApJ, 517, L109
\reference Soderberg, A.~M. \& Ramirez-Ruiz, E. 2002; astro-ph/0210524
\end{references}
\end{document}